\documentclass[superscriptaddress,twocolumn]{revtex4-1}

\usepackage{amsmath}
\usepackage{amssymb}
\usepackage{amsxtra,color}
\usepackage{latexsym}
\usepackage{graphicx}
\usepackage{natbib}
\usepackage[T1]{fontenc}
\usepackage[utf8]{inputenc}
\usepackage{MnSymbol}	

\usepackage{dsfont}
\usepackage{subfigure}
\usepackage{color}
\usepackage{tikz}
\usepackage{multirow}

\makeatletter
\newcommand*{\rom}[1]{\expandafter\@slowromancap\romannumeral #1@}
\makeatother

\newcommand{\be}{\begin{equation}}
\newcommand{\ee}{\end{equation}}

\newcommand{\braket}[2]{\langle \: #1 \: | \: #2 \: \rangle}
\newcommand{\braOket}[3]{\langle \: #1 \: | \: #2 \:| \: #3 \: \rangle}

\newcommand{\intsum}[1]{\sum_{#1} \! \! \! \! \! \! \! \! \! \int }

\newcommand{\E}{{E}}

\newcommand{\delo}{\delta\omega}
\newcommand{\Delo}{\Delta\omega}

\newcommand{\tauSAP}{\tau_{1}}
\newcommand{\tauAPT}{\tau_{32}}

\newcommand{\pv}{\mathrm{p.v.}}

\renewcommand{\Im}{\mathrm{Im}}
\begin{document}

\title{Attosecond photoionization dynamics with stimulated core--valence transitions}
\author{Jhih-An \surname{You}}
\affiliation{Center for Free-Electron Laser Science, Luruper Chaussee 149, 22761 Hamburg, Germany}
\affiliation{Max Planck Institute for the Physics of Complex Systems, Noethnitzerstr. 38, 01187 Dresden}
\affiliation{Max Planck Institute for the Structure and Dynamics of Matter, Luruper Chaussee 149, 22761 Hamburg, Germany}
\author{Nina \surname{Rohringer}}
\affiliation{Center for Free-Electron Laser Science, Luruper Chaussee 149, 22761 Hamburg, Germany}
\affiliation{Max Planck Institute for the Physics of Complex Systems, Noethnitzerstr. 38, 01187 Dresden}
\affiliation{Max Planck Institute for the Structure and Dynamics of Matter, Luruper Chaussee 149, 22761 Hamburg, Germany}
\author{Jan~Marcus \surname{Dahlstr\"om}}
\affiliation{Center for Free-Electron Laser Science, Luruper Chaussee 149, 22761 Hamburg, Germany}
\affiliation{Max Planck Institute for the Physics of Complex Systems, Noethnitzerstr. 38, 01187 Dresden}
\affiliation{Department of Physics, Stockholm University, AlbaNova University Center, SE-106 91 Stockholm, Sweden}
\email{nina.rohringer@mpsd.mpg.de \\ marcus.dahlstrom@fysik.su.se}

\begin{abstract}
We investigate ionization of neon atoms by an isolated attosecond pump pulse in the presence of two coherent extreme ultraviolet or x-ray probe fields. The probe fields are tuned to a core--valence transition in the residual ion and induce spectral shearing of the photoelectron distributions. We show that the photoelectron--ion coincidence signal contains an interference pattern that depends on the temporal structure of the attosecond pump pulse and the stimulated core--valence transition. Many-body perturbation theory is used to compute ``atomic response times'' for the processes and we find strikingly different behavior for stimulation to the outer-core hole ($2p \leftrightarrow 2s$) and stimulation to the inner-core hole  ($2p \leftrightarrow 1s$). The response time of the inner-core transition is found to be comparable to that of state-of-the-art laser-based characterization techniques for attosecond pulses. 
\end{abstract}

\maketitle 


\section{Introduction}
%
Atoms and molecules are today routinely probed and controlled on the atomic time scale 
in various branches of attophysics \cite{krausz:rmp:09}. 
Tailored laser fields are used to control electron trajectories 
and to probe high-order harmonic generation (HHG) \cite{ShafirNature2012,KimNP2013}.
The combination of phase-locked attosecond (as) extreme ultraviolet (XUV) pulses and femtosecond (fs) infrared (IR) laser fields 
has found numerous applications: 
The IR-field can act as an intense probe to 
break chemical bonds \cite{CalegariScience2014} or map time into frequency space by the so-called attosecond streaking techniques \cite{HentschelNature2001,ItataniPRL2002,KeinbergerNature2004,SchultzeScience2010}. 
Alternatively, the XUV pulse can act as a probe 
to study electron/hole dynamics by transient absorption techniques \cite{PhysRevLett.98.143601,GouliemakisNature2010,ott_reconstruction_2014}.
%
%
%
XUV and x-ray free-electron lasers (FELs) \cite{ackermannw._operation_2007, emmap._first_2010, ishikawa_compact_2012, allariae._highly_2012} are accelerator based sources that provide pulses of fs duration with peak powers reaching the gigawatt (GW) range. 
The capabilities of these new sources are rapidly evolving, including the demonstration of wavelength-tunable pulse pairs \cite{ PhysRevLett.110.134801, allaria_two-colour_2013, PhysRevLett.113.254801} 
and spectral--temporal pulse shaping by seeded FELs \cite{GauthierPRL2015}, 
with possible production of GW--as pulses
\cite{PhysRevLett.92.224801, PhysRevSTAB.8.050704,  PhysRevLett.114.244801, PhysRevLett.113.024801}.
Diverse fs pump--probe schemes at FELs led to groundbreaking experiments in chemical reaction dynamics \cite{Erk18072014, mcfarland_ultrafast_2014,wernet_orbital-specific_2015, Barends10092015} and the extension  of these techniques to the as timescale could open new avenues for the observation and control of electron dynamics. 
Attosecond pulses have mainly been characterized using streak-camera techniques 
\cite{HentschelNature2001,ItataniPRL2002,KeinbergerNature2004}, 
or interferometric techniques \cite{PaulScience2001,ChiniOE2010,ChiniNaturePhotonics2014}, 
where photoelectrons are treated as ``replicas'' of the corresponding attosecond pulses that are controlled by a phase-locked optical laser field. 
The implementation of such techniques at FELs is cumbersome due to challenges of synchronization of FEL pulses to optical lasers \cite{schulz_femtosecond_2015
,grgurasi._ultrafast_2012}.
Furthermore, recent difference measurements 
of photoelectrons resulting from different atomic initial states
have evidenced that such XUV--IR schemes, which rely on laser-driven photoelectron dynamics, suffer from uncertainties on the order of tens of attoseconds \cite{SchultzeScience2010,KlunderPRL2011}. 
%
%

\begin{figure}
	\includegraphics[width=\linewidth]{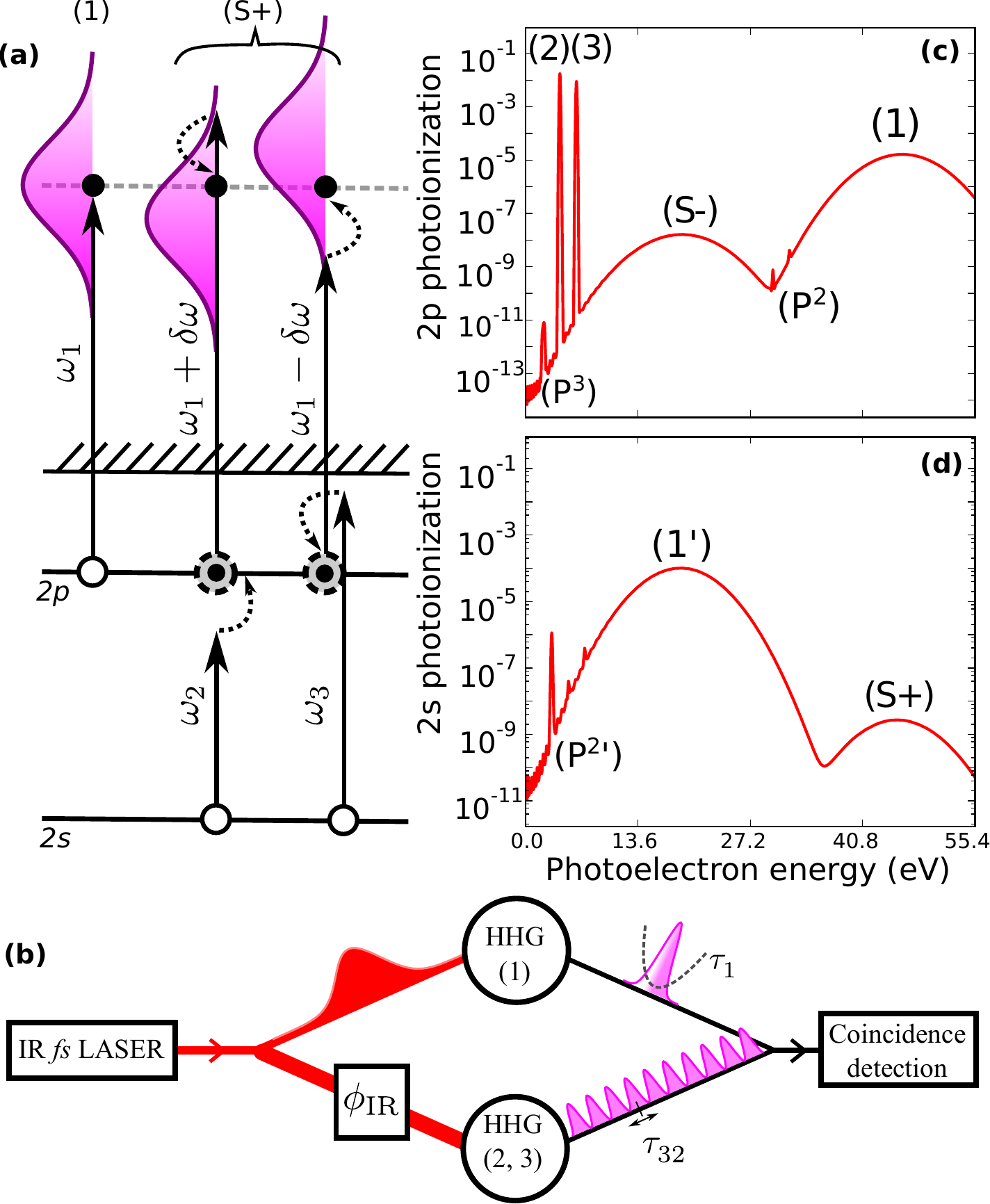}
	\caption{
\label{fig1} 
(Color online)
(a) Photoionization processes in neon: 
(1) photoelectron replica of attosecond pulse by one-photon absorption from $2p$ state; 
and (S+) non-sequential two-photon processes generating spectrally sheared replicas 
by stimulated ionic transitions with a final hole in the $2s$ state. 
(b) Sketch of the proposed experiment where an IR laser field is split into two parts 
for generation of the XUV pump (1) and XUV probe (2,3) fields by HHG. 
Photoionization of neon atoms is then studied with electron--ion coincidence detection.   
(c) Photoelectron spectrum with residual hole in $2p$ state; (d) photoelectron spectrum with residual hole in $2s$ state, 
both computed by 1D-TDCIS (within an independent-particle model). 
See main text for the labeled spectroscopic structures. 
}	
\end{figure}

%
%
This article explores the possibilities to exert control of a photoelectron without laser--electron continuum interaction. To this end, we study a process where an isolated XUV--as pulse ejects an electron from neon, while a pair of XUV (or x-ray) probe fields drives a transition in the ion. The probe fields are tuned to predominately stimulate hole transitions rather than electron continuum dynamics. A small red and blue detuning of the probe fields relative to the resonant hole transition is used to induce spectral shearing of the photoelectron replica to lower and higher kinetic energies, respectively. The shearing process is illustrated under pathway (S+) in Fig.~\ref{fig1}~(a) relative to (1), the unshifted one-photon ionization process. This opens up for a novel control scheme of the final electron energy by {\it indirect} interaction with the probe fields via the hole in the remaining ion. Interestingly, we will show that correlation between the electron and the hole is not required to explain the mechanism and that the process can be described within the independent particle approximation. 

The paper is organized as follows. 
In Sec.~\ref{sec:method} we introduce the proposed scheme and outline our primary theoretical method for the study, which is based on a one-dimensional model of the neon atom. 
In Sec.~\ref{sec:results} we present our main results for the time-dependent one-dimensional model [Sec.~\ref{sec:results-a}] and give an interpretation of the result using three-dimensional perturbation theory [Sec.~\ref{sec:results-b}]. In Sec.~\ref{sec:discussion} we discuss our results and give an outlook for experimental realization of the scheme. Finally, in Sec.~\ref{sec:conclusions} we present our conclusions.

\section{Method}
\label{sec:method}
As illustrated in Fig.~\ref{fig1}~(a), we consider photoionization by an isolated XUV--as pulse in the presence of two coherent XUV (or x-ray) probe fields. 
The complex XUV amplitude of the incoming field on the neon atom written as (atomic units, $\hbar=e=m=4\pi\epsilon_0=1$, are used unless otherwise stated) 
\begin{align}
E(\omega)&=E_1(\omega)+E_2\delta(\omega-\omega_2)+E_3\delta(\omega-\omega_3),
\label{Efieldexpl}
\end{align}
where the pump amplitude is $E_1(\omega)=|E_1(\omega)|\exp[i\phi_1(\omega)]$ with a central frequency $\omega_1$ and a spectral bandwidth $\Delo_1$. 
The peak intensity of the pump pulse is set to $I_1=7\times10^{12}\,$W/cm$^2\,$ with a Fourier-limited pulse duration of $244\,$as.    
The group delay 
\be 
\tauSAP(\omega)=\frac{d\phi_1}{d\omega}, 
\label{taugd}
\ee
describes the arrival time of a particular frequency component $\omega$ of the pump pulse at the target. We consider probe frequencies $\omega_{f=2,3}$, that are symmetrically red and blue shifted relative to the $2p\leftrightarrow2s$ hole transition 
\begin{align}
\omega_2&=\epsilon_{2p}-\epsilon_{2s}-\delo \nonumber \\
\omega_3&=\epsilon_{2p}-\epsilon_{2s}+\delo, \nonumber
\end{align}
where the outer hole energy is $\epsilon_{2p}=-21.56\,$eV and the inner hole energy is $\epsilon_{2s}=-48.47\,$eV \cite{NIST_ASD}. 
The peak intensity of the probe fields is set to $I_f=3.5\times10^{12}\,$W/cm$^2$. 
The detuning is supposed small compared to the bandwidth of the pump field, 
$\delo\ll\Delo_1$, 
and the probe fields are quasi-monochromatic with a bandwidth much smaller than the detuning, 
$\Delta\omega_{2,3}\ll\delo$, as indicated by the delta functions in Eq.~(\ref{Efieldexpl}). 
The group delay of the probe fields is defined as $\tauAPT=(\phi_3-\phi_2)/2\delo$ 
using the spectral phase of the probe fields, $\phi_{2,3}=\arg\{E_{2,3}\}$. 
In Fig.~\ref{fig1}~(b) we propose an experimental setup where the pump (1) is generated by HHG from an ultra-short IR pulse and the probe fields (2,3) are odd HHG harmonics from a longer IR pulse. The group delay of the probe fields is then locked to the phase of the IR laser field due to the non-linear HHG process, $\tauAPT\propto\phi_\mathrm{IR}$, and can be accurately controlled by an IR laser-delay stage. 
Implementation of this probe technique at FEL sources would require two-color pulse pairs \cite{ PhysRevLett.110.134801, allaria_two-colour_2013, PhysRevLett.113.254801} and accurate temporal and phase control \cite{GauthierPRL2015}. 

As depicted in Fig.\ref{fig1}~(a), the proposed probe process can be cast in terms of single-particle transitions. An appropriate numerical method is therefore time-dependent configuration interaction singles (TDCIS) \cite{RohringerPRA2006}. For the calculation of the photoelectron spectrum we implemented the coupled surface flux method \cite{TaoNJP2012}, similar to the approach used in Ref.~\cite{Karamatskou}. To discuss the  basic process, we opt for a one-dimensional (1D) description of the considered process. 
In Fig.~\ref{fig1}~(c)~and~(d) we show simulated ionic channel resolved photoelectron spectra for a 1D model of the neon atom. 
Surprisingly, we have found that it was important to account for stimulated ion dynamics by the probe fields after the electron has escaped the inner region, as explained in Appendix A. Electron--electron correlation effects do not influence the ionization process significantly and this allows us to further approximate the system by an independent-particle model, where the correlation coupling terms in TDCIS are neglected. 
This makes the interpretation of the numerical results more tractable but also helps to speed-up the computational time. 
%

\section{Results}
\label{sec:results}
In Sec.~\ref{sec:results-a} we present our numerical results for the time-dependent 1D model of neon. 
In order to interpret our results and to make quantitative estimates we then turn to perturbation theory in Sec.~\ref{sec:results-b}, 
where we first consider the stimulated $2p\leftrightarrow 2s$ (XUV) transition in Sec.~\ref{sec:2p2s} 
and then the $2p\leftrightarrow 1s$ (x-ray) transition in Sec.~\ref{sec:2p1s}.  

\subsection{Time-dependent 1D model}
\label{sec:results-a}
Photoelectrons leaving the residual ion with a hole in the $2p$ state [Fig.~\ref{fig1}~(c)] exhibit one broad peak (1) due to absorption of one pump photon (with $\omega_1=68$\,eV and $\Delta\omega_1=7.5$\,eV) 
and two narrow peaks (2,\,3) due to absorption of either probe field (with $\omega_{2,3}=26.9\mp1$\,eV and $\Delta\omega_{2,3}=0.125$\,eV). 
The broad peak (S-) is due to absorption of one pump photon and stimulated emission of one probe photon.
Weaker peaks labeled with (P$^2$) and (P$^3$) denote two and three probe-photon processes, respectively.
Photoelectrons leaving the residual ion with a hole in the $2s$-state [Fig.~\ref{fig1}~(d)] exhibit a peak (1') from the pump field and a peak (S+) due to absorption of one pump photon and one probe photon. 

\begin{figure}
	\includegraphics[trim=1.5cm 0.5cm 1.5cm 0cm,width=\linewidth]{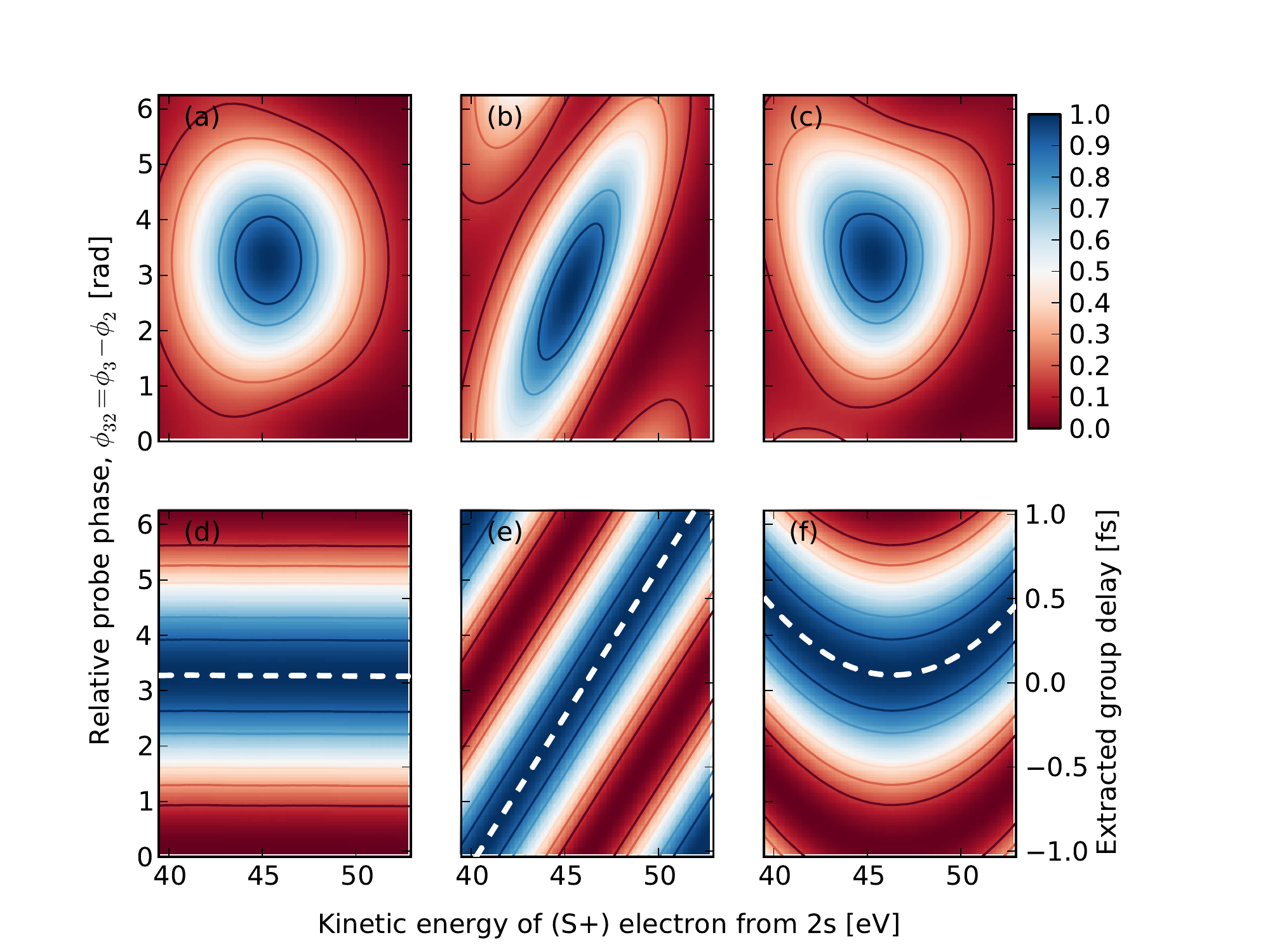}
	\caption{\label{fig2} 
(a)--(c) (Color online) Normalized photoelectron distribution of the (S+) peak in Fig.~\ref{fig1}~(d), as a function of phase difference between probe fields, $\phi_{32}$. Data is computed using 1D-TDCIS (within an independent-particle model). (a) Fourier limited pump pulse; (b) linear chirp; and (c) quadratic chirp of pump pulse. 
(d)--(f) same as (a)--(c) but with normalization at each individual kinetic energy. 
The left vertical axis labels the relative phase in radians, while the right axis labels the extracted group delay of the attosecond pulse in femtoseconds defined in Eq.~(\ref{taugd}), shown by the dashed white curve.
}	 
\end{figure}

In Fig.~\ref{fig2} we show the behavior of the (S+) structure as a function of relative probe phase $\phi_{32}=\phi_{3}-\phi_{2}$, for (a) a Fourier limited pump pulse, $\phi_1(\omega)=0$; (b)  a quadratic phase dependence, $\phi_1(\omega)=\alpha(\omega-\omega_1)^2$ with $\alpha=100$; and (c) a cubic phase dependence, $\phi_1(\omega)=\beta(\omega-\omega_1)^3$ with $\beta=100$. 
%
%
\begin{figure}
	\includegraphics[trim=1.5cm 0.5cm 1.5cm 0cm,width=\linewidth]{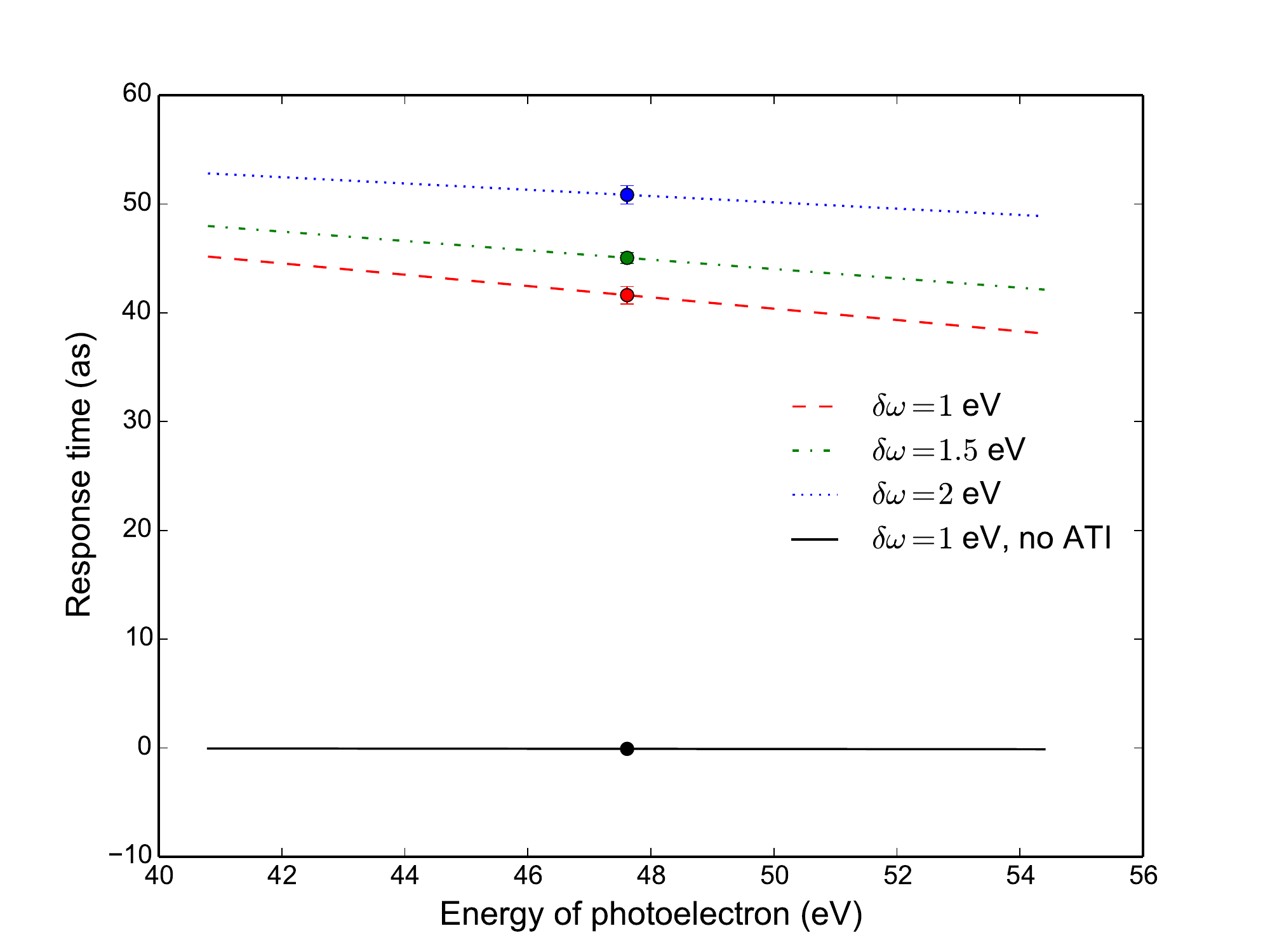}
	\caption{\label{fig3pra} 
(Color online) Detailed study of the ``response time'' of the (S+) peak for three different detuning of the probe field given an unchirped pump pulse ($\alpha=\beta=0$). In accordance with Eq.~(\ref{Wpb}), the response time $\tau_{pb}$, is extracted by making a cosine fit to the phase-dependent oscillations of the photoelectron probability, e.g. the modulations shown in Fig.~\ref{fig2}~(d) (where $\delo=1\,$eV). The raw data (not shown) has been fitted to a line in order to extract the $\alpha$-parameter of the pump pulse. Standard deviation of the linear fit is indicated by the error bars. The data was computed by the 1D time-dependent independent-particle model discussed in Appendix A. 
}	 
\end{figure}
%
%
\begin{table*}[ht]
	\centering        
	\begin{tabular}{| c | c | c | c | c | c | c | c |}        
		\hline 
		\multirow{2}{*}{$\alpha$} & \multirow{2}{*}{$\delta \omega $ (eV)} & \multirow{2}{*}{$\Delta\omega_1$ (eV)} & \multirow{2}{*}{$\Delta\omega_f $ (eV)} & \multirow{2}{*}{$\tilde{\alpha}$} & \multicolumn{3}{c|}{Numerical} \\[0.5ex]
		\cline{6-8}
		& & & & & with ATI & Without ATI & Difference  \\
		\hline
		-100 & 1 & 7.5 & 0.0625 & -99.79  & -100.127  & -99.85 & -0.277\\
		100 & 1 & 7.5 & 0.0625 & 99.79  & 99.55  & 99.86 & -0.31\\
		100 & 1 & 7.5 & 0.125 & 99.5 & 99.058 & 99.39 & -0.332\\ 
		10 & 1 & 7.5 & 0.125 & 9.953 & 9.646 & 9.943  & -0.297\\ 
		100 & 0.5 & 7.5 & 0.125 & 97.69 & 97.149 & 97.57 & -0.421\\ 
              [1ex] 
		\hline            
	\end{tabular}
		\caption{Comparison of the retrieved $\alpha$ from numerical calculations and analytical estimate ($\tilde\alpha$).}
	\label{table:nonlin}  
\end{table*}
In Fig.~\ref{fig2}~(a) the (S+) peak vanishes for $\phi_{32}\approx0$, while the peak is maximized for $\phi_{32}\approx\pi$. 
As will be derived in Sec.~\ref{sec:results-b}, this is due to a relative $\pi$-shift between the two-photon paths that have positive and negative detuning with respect to the hole resonance, respectively.  
In Fig.~\ref{fig2}~(d)--(f) we show more clearly the $\phi_{32}$-dependence of the (S+) structure by dividing every phase-dependent curve, at a fixed energy of Fig.~2.~(a)-(c), by its maximal value. In analogy with spectral shearing interferometry, the modulations (indicated by white dashed curves) are expected to depend on the group-delays of the attosecond pump pulse with (d) constant value (e) linear chirp and (f) quadratic chirp, respectively. By direct fit to the unchirped case ($\alpha=\beta=0$) we obtain an extracted  $\alpha$-value of $-0.288$, corresponding to a drift of $-0.5\,$as/eV. 
Where does this ``response time'' come from?
%
%

Further, our simulations show that the (S-) peak [Fig.~\ref{fig1}~(c)] has a similar $\phi_{32}$-dependence as (S+). The one-photon peaks (1) and (1') also modulate with $\phi_{32}$, but the variation is opposite to that of (S+) and (S-). Physically, this is because the probe fields are effectively shifting the ionic channel of the photoelectrons, e.g. 
by redistribution of population from peak (1) to peak (S+). The total photoelectron spectrum, unresolved for the residual ionic state, does not show clear $\phi_{23}$-dependence. Unfortunately, this makes the experimental measurement of the (S+) modulations challenging, because it must rely on coincidence detection, as we will discuss in Sec.~\ref{sec:discussion}. 
%
%

In Fig.~\ref{fig3pra} we show the ``response time'' for the case of an unchirped pump pulse, i.e. by zooming in on the dashed curve in Fig.~\ref{fig2}~(d), for three different symmetric detuning of the probe fields, $\delo=1,\,1.5$ and $2\,$eV. All detuning show qualitatively the same result with a response time in the range 40--55 attoseconds. All curves exhibit a negative slope with extracted $\alpha$-values for detuning $1\,$eV, $1.5\,$eV and $2\,$eV equal to $-0.288$, $-0.239$ and $-0.160$, respectively. In order to interpret this behavior we have additionally performed simulations where the photoelectron does not interact with the probe field. In Fig.~\ref{fig3pra} we label this result as ``no ATI'' (no above-threshold ionization) because the electron can not absorb probe photons after it has been ejected within this model (see also Appendix A). Interestingly, the extracted $\alpha$-parameter is $-0.00294$, which is much closer to the expected zero value. This shows that XUV driven electron continuum dynamics must be responsible for the finite response time. 
The discrepancy in recovering the parameters of the pump pulse is attributed to an ``atomic response time'' 
and a derivation of how the group delay $\tauSAP(\omega)$, is mapped to the (S+) structure will be given 
by perturbation theory in the next section.  
In the case of a chirped pump pulse, e.g. $\alpha\ne0$ case, the finite duration of the probe fields will also affect the extracted values. In the simplified case of Gaussian pulses (with linear chirp given by $\alpha$) and by considering the ``no ATI case'', we have found that this effect can be approximated as 
\be
\tilde\alpha=\alpha\left(1-\frac{2\Delta\omega_f^2}{\Delta\omega_1^2}-\frac{\Delta\omega_f^2}{4\ln(2)\delo^2}\right), 
\ee
where $\tilde\alpha$ is the extracted value for the $\alpha$-parameter of the pulse. 
The bandwidths of the pump and probe pulses are labeled as $\Delta\omega_1$ and $\Delta\omega_{f={2,3}}$, respectively,  
and defined as the full-width at half-maximum (FWHM), $E_i(\omega)\sim\exp[-2\ln (2)(\omega-\omega_i)^2/\Delta\omega_i^2]$. In table~\ref{table:nonlin} we show the reasonable agreement between this simple analytical estimate ($\tilde\alpha$) and the numerical simulations (without ATI). Further, we note that the difference between the case with ATI and without ATI (right-most column) with $\delo=1\,$eV is approximately $-0.3$ in agreement our finding for the $\alpha=0$ case. This indicates that the electron contribution does not depend strongly on the chirp of the pump pulse. Finally, the difference between the numerical simulations with $\alpha=100$ and $\delo=0.5\,eV$ (in table~\ref{table:nonlin}) shows a larger negative slope than the $\delo=1\,$eV case, in agreement with the trend found for the $\alpha=0$ case (in Fig.~\ref{fig3pra}). 
In the following we will not focus on these detailed pulse convolution effects, that occur due to the finite bandwidth of the probe fields $\Delta\omega_f$, but rather explain the fundamental reason for why electron continuum dynamics leads to a non-zero response time.        


\subsection{Time-independent perturbation theory}
\label{sec:results-b}
In order to better understand the physical mechanism of the atomic response time we now turn to perturbation theory. The dominant complex amplitudes that give rise to $\phi_{32}$-dependent modulations of the (S+) structure are given by 
\begin{align}
S_{pb,f}=
\frac{1}{i}\E_1(\omega_{pb}-\omega_f)\E_fM_{pb,f}, \, \, \, f=2,3
\label{Spbf}
\end{align} 
corresponding to absorption of one photon from the pump field with energy $\omega_1'=\omega_{pb}-\omega_f$, followed by one from either probe field with energy $\omega_{f}$. 
\subsubsection{Stimulation of $2p\rightarrow 2s$ hole transition}
\label{sec:2p2s} 
In Eq.~(\ref{Spbf}), the two-photon matrix element $M_{pb,f}=M_{pb}(\omega_{pb}-\omega_f,\omega_f)$, describes a transition from the ground state to a final state with one electron in the continuum state $p=ks,kd$ (with energy $\epsilon_p>0$) and a hole in the atomic orbital $b=2s$ ($\epsilon_b<0$). 
	Energy conservation is imposed as $\omega_{pb}=\epsilon_p-\epsilon_b=\omega_1'+\omega_f$.   
The probability density for electrons in the (S+) structure is computed as the square the two complex amplitudes with $f=2,3$ leading to an  interference pattern over $\phi_{32}\propto\tauAPT$  
\begin{align}
W_{pb}&
\approx|S_{pb,2}+S_{pb,3}|^2 
\nonumber \\
=&|A_{pb}|-|B_{pb}|\cos\left[2\delo\left(\tauSAP-\tauAPT+\tau_{pb}\right)\right],
\label{Wpb}
\end{align} 
where $|A_{pb}|$ is the incoherent sum of the transition strengths, while $|B_{pb}|$ relates to the cross term of the amplitudes.
We note that the group delays of the pump $\tauSAP$ and the probe fields $\tauAPT$ enter with opposite signs in Eq.~(\ref{Wpb}), 
which must be the case because if the pump field is delayed by a certain amount the probe field must also be delayed by the same amount to recover same outcome. 
Besides the group-velocity delays 
$\tauSAP$ and $\tauAPT$, 
the interference pattern is delayed by 
\begin{align}
\tau_{pb}=[\arg(M_{pb,2}M_{pb,3}^*)-\pi]/2\delo,
\label{taupb}
\end{align}
which depends on the phase difference between the two-photon matrix elements, and can be interpreted as an atomic response time for creating the (S+) peak.  
Because it is convenient to define the response time as a small number we have included a $-\pi$ inside the square bracket in Eq.~(\ref{taupb}) to  remove the relative $\pi$-shift between the two-photon transitions due to the resonance. It follows from this definition that there is a minus sign on the cosine in Eq.~(\ref{Wpb}).  
While $\tau_{pb}$ may be regarded as a limiting factor for determining the unknown $\tau_1$, explaining the modest error in the extracted $\alpha$-value above for the 1D model, the response time is an interesting quantity to study further, as it contains information about the stimulated core--valence transition, in particular, it contains information about the phase difference of the two-photon (XUV--XUV) matrix elements.       

In order to make a quantitative estimate of $\tau_{pb}$ we now turn to many-body perturbation theory to describe neon in 3D and include correlation effects.
Following Ref.~\cite{DahlstromPRA2012}, our calculations are based on single-particle states that are expanded on a spherical basis $\phi_i(\mathbf{r})=R_{n_i,l_i}(r)Y_{l_i,m_i}(\mathbf{\hat r})$. The radial wavefunctions $R_i(r)$, are eigenstates to the restricted Hartree-Fock (HF) equation for occupied states, while the unoccupied (virtual) states are additionally attracted by an effective spherical potential to model the long-range Coulomb interaction between electron and ion.
%
\begin{figure}
	\includegraphics[width=\linewidth]{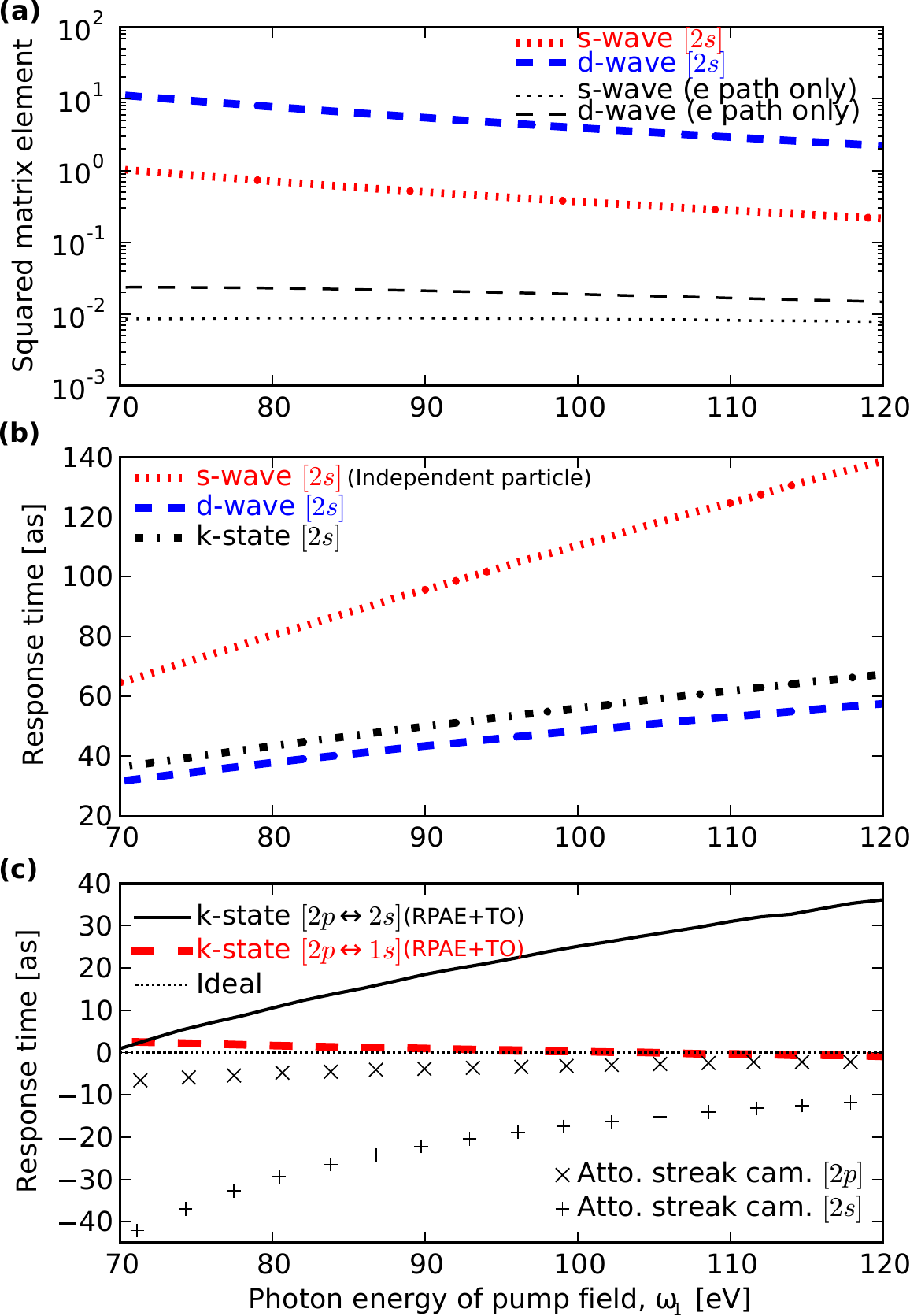}
	\caption{\label{fig3}
(a) (Color online) Squared matrix elements $|M_{pb,f}|^2$, for final s-wave (dotted) and d-wave (dashed), including both stimulated hole and electron paths in bold line and only electron path in thin line.  
(b) Response time for photoemission along polarization axis, s-wave and d-wave. Data presented in (a) and (b) are computed by a 3D independent-particle model of neon. 
(c) Response time for photoemission along the polarization axis (within a correlated model including both time-orderings) for the stimulated valence hole ($2p\rightarrow2s$) and core hole ($2p\rightarrow1s$) transitions. The streak-camera delay from initial 2p (2s) state \cite{DahlstromPRA2012} is shown for reference. 
}	
\end{figure}
\begin{figure}
	\includegraphics[width=\linewidth]{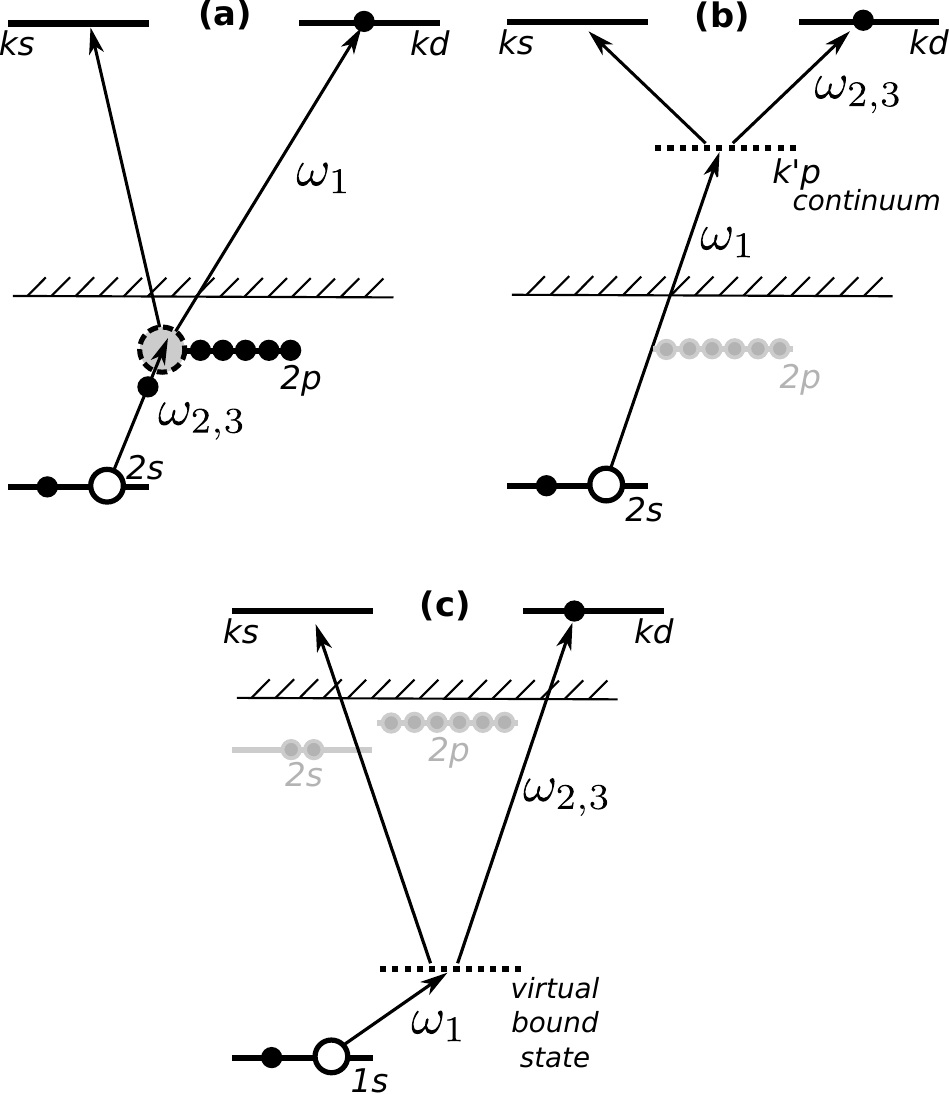}
	\caption{\label{fig5pra}
(a) Two-photon diagram for stimulated hole transition, $2p \rightarrow 2s$, after photoemission from outer state, $2p\rightarrow ks,kd$.
(b) Two-photon diagram for stimulated electron continuum transition from an inner valence state, $2s\rightarrow k'p\rightarrow ks,kd$. 
(c) Two-photon diagram for stimulated virtual electron transition from a core state, $1s\rightarrow n'p\rightarrow ks,kd$.
}	
\end{figure}
%
%
The two-photon matrix elements are separated into two terms using second quantization  
\be
M_{pb,f} \approx M_{pb,f}^\mathrm{(hole)} + M_{pb,f}^\mathrm{(elec.)}, 
\label{Msum}
\ee
because the probe field can either stimulate a hole transition 
or a continuum electron transition, 
as illustrated in Fig.~\ref{fig5pra}~(a) and (b), respectively.
The stimulated hole term
\be
M_{pb,f}^\mathrm{(hole)}=
\sum_{a'}\frac{
z_{a'b}z_{pa'}
}
{(\omega_f-\epsilon_{a'}+\epsilon_b)}, 
\label{hh}
\ee  
describes a dipole transition of an electron from any occupied single-particle state $\phi_{a'}$, to the final electron state $\phi_p$, followed by a dipole transition of the hole to the final state $\phi_b$. The radial orbitals are chosen to be real, which implies a real transition to a given final partial wave state, $\phi_p=ks,kd$ for $\phi_b=2s$. 
In contrast, the stimulated electron term is complex 
\begin{align}
M_{pb,f}^\mathrm{(elec.)}
=&
\lim_{\xi\rightarrow 0^+}
\intsum{p'}
\frac{
z_{pp'}z_{p'b}
}
{(\omega_1'-\epsilon_{p'}+\epsilon_{b}+i\xi)}
\nonumber \\ 
=&
\pv\intsum{p'}
\frac{
z_{pp'}z_{p'b}
}
{(\omega_1'-\epsilon_{p'}+\epsilon_{b})}  
-i \pi 
z_{pr}z_{rb}
,
\label{cc}
\end{align}
and it describes an initial dipole interaction
that excites an electron from the occupied state $\phi_b=2s$, to the unoccupied states, $\phi_{p'}=n'p$ and $k'p$. 
The second dipole interaction
then stimulates an electron transition in the continuum ($\omega_1'+\epsilon_b>0$ as shown in Fig.~\ref{fig5pra}~(c)) to the final state $\phi_p=ks,kd$. 
%
In Eq.~(\ref{cc}) we 
write the matrix element as real non-resonant contributions (principal-value sum--integrals over $p'=k'p$) and an imaginary resonant contribution (via the intermediate state $\phi_r=k_rp$ with $\epsilon_r=\epsilon_b+\omega_1'$). 
If the probe field is in the IR range the stimulated electron transition is a good approximation for the total two-photon matrix element $M_{pb,f}\approx M_{pb,f}^\mathrm{(elec.)}$, but this is not the case for the processes studied here with nearly resonant XUV transitions.   
In Fig.~\ref{fig3}~(a) we show that stimulated electron contributions $|M_{pb,f}^\mathrm{(elec.)}|^2$ are 2--3 orders of magnitude smaller than the total contributions $|M_{pb,f}|^2$  that are dominated by the strong resonant coupling in the residual ion.  
%
%
Here, the total matrix element is better approximated as 
a hole transition plus a small imaginary electron transition 
$
M_{pb,f}\approx M_{pb,f}^\mathrm{(hole)}+i\Im M_{pb,f}^\mathrm{(elec.)}.
$
If we note that $M_{pb,2}^\mathrm{(hole)}=-M_{pb,3}^\mathrm{(hole)}$ and assume that $M_{pb,2}^\mathrm{(elec.)}\approx M_{pb,3}^\mathrm{(elec.)}$ 
then
\begin{align}
\tau_{pb}\approx&
-\frac{\arg\left[M_{pb,3}\right]}{\delo} \nonumber \\
=&\frac{1}{\delo}
\arctan\left[
\frac{\pi z_{pr}z_{rb}}{z_{ab}z_{pa}/\delo}
\right]
\approx \frac{\pi z_{pr}z_{rb}}{z_{ab}z_{pa}},
\label{taupbapprox}
\end{align}
where $z_{ij}>0$ are dipole matrix elements between between the real single particle states $\phi_i$ and $\phi_j$.
The last step in Eq.~(\ref{taupbapprox}) is valid for small detuning, $\delo \ll |z_{ab}z_{pa}/\pi z_{pr}z_{rb}|$. 
Interestingly, Eq.~(\ref{taupbapprox}) shows that $\tau_{pb}$ does not depend strongly on $\delo$ but rather gives direct information about the ratio between dipole matrix elements of the stimulated electron and hole transitions. 
In other words, decreasing the detuning to enhance further stimulate the hole transitions will not alter the response time. 
This has been verified by many-body perturbation theory for $\delo=1$ and $1.5\,$ eV, 
where we found that the response time changed by less than one attosecond. 

%
In Fig.~\ref{fig3}~(b) we present $\tau_{pb}$ from Eq.~(\ref{taupb}) using the 3D independent particle model given by Eq.~(\ref{Msum}). 
A positive linear drift is found on both final partial waves, $ks$ and $kd$, which we attribute to an increasing relative contribution from the resonant  electron path, i.e. the numerator on the right side of Eq.~(\ref{taupbapprox}).   
We also show $\tau_{pb}$ for photoelectrons with momentum $\mathbf{k}=k\mathbf{\hat z}$ along the polarization axis, computed by the complex final state 
\begin{align} 
\phi_{\mathbf{k}}(\mathbf{r})\propto\sum_{L,M}i^{L}e^{-i\eta_{L}(k)}Y_{L,M}^*(\mathbf{\hat k})R_{k,L}(r)Y_{L,M}(\mathbf{\hat r}),
\end{align}
where $\eta_L(k)$ are scattering phases of the real radial functions, $R_{k,L}(r)$. The angle-resolved emissions has a linear drift of $0.634$ as/eV, quite close to the dominant $d$-wave. Over a large energy range, from 65 to 120 eV, the deviation from this linear fit is less than one attosecond.  
The response times of the 3D calculation are in qualitative agreement with those of the 1D case, with a delay on the order of tens of attoseconds. 
Surprisingly, we find that the slopes of the response times are different in the 1D and 3D case. 
It remains an open question if this discrepancy is entirely due to pulse convolution effects or if the different electronic structure between 1D and 3D plays a role. In order to answer this question it would be beneficial to perform 3D TDCIS calculations \cite{Karamatskou}, but this remains an endeavor beyond the scope of the present paper. 

Next, we add correlation effects by implementing the random phase approximation with exchange (RPAE) on the first dipole interaction \cite{DahlstromPRA2012}, which increases the linear slope to $0.733\,$as/eV (not shown). Including also the non-resonant, reversed time-order processes (TO), where the probe photon is absorbed before the pump photon changes the slope marginally to $0.724\,$as/eV. Although the contribution from the reversed-TO is rather small, we stress that the $\delo$-dependence reported for the 1D model in Fig.~\ref{fig3pra}, can not be explained without taking this effect into account. 

Finally, we note that the ratio of the two-photon (pump+probe) and one-photon (pump) transition rates 
\be
R\approx \left|\frac{E_f z_{ba}}{\delo}\right|^2,
\ee 
scales inversely with the squared detuning of the probe field. This implies a boost of the two-photon transition by tuning the probe fields closer to the resonance. Assuming $\delo=1\,$eV $=0.0358\,$ au, $z_{ab}\approx 1\,$ au and $R=1\%$, we estimate the required probe field intensity is $I_f|_{R=1\%}=7.4\times10^{-5}\,$au$ = 2.6\times 10^{12}\,$W/cm$^2$. In the case studied here with one short pump pulse and two long probe fields it is only the temporal overlap of the two pulses that will contribute to the two-photon transition. Using the time-dependent model, discussed in Sec.~\ref{sec:results-a}, we have verified that the probability for the two-photon transition does not depend on the duration of the probe fields, but rather on the instantaneous intensity and detuning of the probe fields. We refer the interested reader to Ref.\cite{Jimenez2015} for an insightful discussion about pump--probe schemes on the attosecond time scale.     

\subsubsection{Stimulation of $2p\rightarrow 1s$ hole transition}
\label{sec:2p1s}
The squared two-photon matrix element for the $2p \rightarrow 1s$ stimulated hole transition is roughly two order of magnitude smaller than that of the $2p \rightarrow 2s$, shown in Fig.~\ref{fig3}~(a), but the trend is otherwise similar. This is easy to understand because the dipole coupling from the $2p$ valence state to the $1s$ inner core state is smaller than that of $2p$ to $2s$.    
In Fig.~\ref{fig3}~(c) we compare $\tau_{pb}$, including correlation and both time-orders (RPAE+TO), for  XUV-stimulated outer-core--valence transition ($2p \rightarrow 2s$) and x-ray inner-core--valence transition ($2p \rightarrow 1s$). Evidently, the core transition has a much shorter response time. This can be explained by the fact that the (main) electron path no longer goes through the continuum, but instead on a virtual bound excitation 
\be
M_{pb,f}^\mathrm{(elec.)}=\intsum{n'}\frac{z_{pn'}z_{n'b}}{(\omega_1'-\epsilon_{n'}+\epsilon_b)},
\label{Mel1s}
\ee
where $\omega_1'+\epsilon_b<0$ for $b=1s$, as illustrated in Fig.~\ref{fig5pra}~(c). 
In the present calculation we used the real HF energy for the 1s orbital $\epsilon_{1s}^{(HF)}=-891.70\,$eV, 
which should be valid provided that the pump pulse is sufficiently short.  
Nonetheless, we have also tested to give the $1s$-energy an imaginary part (equal to $0.27\,$eV) to mimic the decay of the core hole, 
but this did not change the response time by more than one attosecond. 

%
%

\section{Discussion}
\label{sec:discussion}
In this paper we have explored a novel idea to perform spectral shearing interferometry of photoelectrons using two coherent XUV (or x-ray) probe fields. Due to the excess or shortage of photon energy for a given ionic transition, the photoelectron will shift up or down in coincidence with the transition in the ion. The idea is closely related to the attosecond streak-camera method \cite{HentschelNature2001,ItataniPRL2002,KeinbergerNature2004} where a strong IR field is used to drive the electron in the continuum and to the PROOF method \cite{ChiniOE2010,LaurentOE2013} where a single IR photon is absorbed or emitted to shear the photoelectron distribution. 
The corresponding atomic response times of the attosecond streak camera are shown for reference in Fig.~\ref{fig3}~(c) \cite{DahlstromPRA2012}. 
As can be observed, the response time from the outer-core method ($2p\rightarrow 2s$ transition) is larger than the response time of the streak camera from the $2p$ state in neon. As we explained, the relatively large response of the outer-core method comes from stimulated continuum electron transitions by the probe field. In contrast, the response time of the inner-core method ($2p\rightarrow 1s$) is found to be comparable to  that of the streak-camera method. In this case the response time of the inner-core method comes from correlation effects and possibly field-convolution effects. In theory, this establishes the newly proposed scheme as an all-XUV or x-ray method for direct group-delay determination of attosecond pulses. Assuming that the attosecond pulse has been readily characterized, e.g. by the streak-camera method, the new method can be used to study the phase difference of two-photon (XUV--XUV or XUV--x-ray) processes.  
However, in order to extract the desired signal, i.e. the $\phi_{32}$-dependent modulations of the (S+) peak in Fig.~\ref{fig2}, we need to study channel-resolved photoelectrons. More precisely, we need to distinguish between electrons from the unexcited ion (with a $2p$ hole) and the excited ion (with a $1s$ or $2s$ hole). In practice, this is a major drawback of the new method because the streak-camera does not require any form of coincidence detection. 
The first coincidence detection schemes combined with attosecond pulses have been reported recently \cite{RanitovicPRL2011,SabbarRSI2014,ManssonNP2014}, but so far no experiments have been reported where the state of the ion has been determined separately from the electron.  
Nonetheless, let us now speculate as to how this type of measurements can be performed in future experiments, inspired by the existing technology, such as reaction microscopes \cite{Ullrich2003} and photoelectron--fluorescence coincidence detection \cite{RubenssonPRL1996}. 
First, for the case of a $1s$ hole, high-energy Auger emission will occur on a femtosecond time scale and efficiently convert the singly charged ion to the doubly charged ion. A reaction microscope can be used to separate the photoelectron and the ion in space, then the ionic charge can be determined by accellerating the ions in an electric field. Since Auger emission is the dominant decay mechanism for the $1s$ hole this Auger-based method is deemed more feasible than fluorescence-based detection. 
In contrast, the decay occurs exclusively by fluorescence for the case of a $2s$ hole and it may appear that the only way to probe the state of the ion would be to detect florescence photons on a nanosecond time scale. However, recent experimental work has shown that it is possible to laser-enable Auger decay of the $2s$ hole by hitting the excited ion with an intense IR laser field \cite{RanitovicPRL2011}. This opens up for both ion acceleration technique  and electron-coincidence detection of the high-energy primary electron and the low-energy Auger electron to determine the state of the ion.  
Clearly, all these ideas are more challenging to implement experimentally than the conventional attosecond streak camera, but we believe that these are technical challenges that can be overcome in the future.  
Finally, we stress that the issue of photoelectrons with the same final energy from different states of the ion is inherent to the broad bandwidth of the pump pulse. If our aim is to study the phase of the two-photon matrix elements it is more efficient to {\it replace} the isolated pump pulse [(1) in Fig.~\ref{fig1}] by an attosecond pulse train that translates to a comb-like photoelectron spectrum with spacing $2\delo$. In this case the (S+) signal then resides on peaks in between the comb-like peaks of the pump field, which means that the $\phi_{32}$-dependence can also be studied without coincidence detection, but only at discrete energy intervals determined by the comb structure. This setup bears great resemblance with and could be used together with the RABBITT method \cite{PaulScience2001}.       

\section{Conclusions}
\label{sec:conclusions}
In conclusion, we have proposed a new type of pump--probe scheme that relies on stimulated core--valence transitions by two narrow-band detuned XUV or x-ray probe fields and a short XUV pump pulse. Here, we applied the method to the characterization of isolated attosecond pulses and we demonstrated the existence of an atomic response time that gives insight into the nature of the stimulated core--valence transitions. 
In particular, for the stimulated $2p\rightarrow 2s$ hole transition in neon, we found that the response time can be approximated by a ratio between electron continuum transitions and the stimulated hole transition.     
In practice, the method relies on coincidence detection of electron and ion, which makes it less efficient than existing techniques based on IR sources for pulse characterization. Nonetheless, the method is a natural candidate for future XUV--XUV experiments on table-top HHG sources and at FEL facilities, as it presents a way to study XUV--XUV/x-ray processes with short pump pulse probed by the sharp frequency bandwidth of the probe fields. 


%
%

\section*{Appendix A}

Time-dependent configuration interaction singles (TDCIS) \cite{RohringerPRA2006} includes the Hartree-Fock ground state $|\Phi_0 \rangle$ and its single excitations $|\Phi_a^p \rangle$ based on the one-particle Fock operator $\hat{H_0}$ and its eigenstate $|\varphi_i\rangle$ with energy $\epsilon_i$.
Generally, indices $a$,$b$,$c$... are used for spatial orbitals that are occupied in $|\Phi_0 \rangle$, for unoccupied (virtual) orbitals indices $p$,$q$,$r$,...are employed, and the indices $i$,$j$,$k$,... are for general orbitals (occupied or unoccupied). 
Spin-orbit interaction is not considered in this work.
The many-body wave packet in the CIS basis is given by
\begin{equation} \label{wave_packet}
|\Psi,t \rangle=\alpha_0(t)|\Phi_0 \rangle+\sum_{p}\sum_{a}\alpha_a^p(t)|\Phi_a^p \rangle,
\end{equation}
with initial conditions $\alpha_0(t_0)=1$ and $\alpha_a^p(t_0)=0$.
To describe the hole dynamics and the corresponding electron wave packet propagating in the real space, 
we introduce time-dependent orbitals that collect all single excitations originating from the occupied orbitals $|\varphi_a\rangle$,
\begin{equation}
|\chi_a(t)\rangle=\sum_{p}\alpha_a^{p}(t)|\varphi_p\rangle.
\end{equation}
For the atomic system interacting with laser field $E(t)$ linearly polarized along the z axis, the TDCIS equations of motion can be written as
\begin{equation}
i\dot{\alpha_0}=-E(t)\sum_a \langle\varphi_a|\hat{z}|\chi_a(t)\rangle
\end{equation}
\begin{align} \label{TDCIS}
i\frac{\partial}{\partial t}|\chi_a(t)\rangle&=(\hat{H}_0-\varepsilon_a)|\chi_a(t)\rangle + 
\sum_{b} \hat{P}\{\hat{K}_{ba}-\hat{J}_{ba}\}|\chi_b(t)\rangle
\nonumber \\ &-E(t)\hat{P}\hat{z}\{\alpha_0 |\varphi_a \rangle +|\chi_a(t) \rangle \}+E(t)\sum_{b}z_{ba}|\chi_{b}(t) \rangle,
\end{align}where $z_{ba}=\langle\varphi_b|z|\varphi_a\rangle$, $\hat{P}$ is the projection operator acting on the subspace composed of the virtual orbitals
\begin{equation}
\hat{P}=\sum_p |\phi_p\rangle\langle\phi_p|=1-\sum_a|\phi_a\rangle\langle\phi_a|, 
\end{equation}
and $\hat{J}_{ba}$ and $\hat{K}_{ba}$ are, respectively, generalized Coulomb and Exchange operators associated with the direct Coulomb matrix elements $v_{pbqa}$ and the exchange matrix elements $v_{pbaq}$:
\begin{equation}
\begin{split}
v_{pbqa}& \equiv  \langle \varphi_p |\hat{J}_{ba}| \varphi_{q} \rangle \\
v_{pbaq}& \equiv  \langle \varphi_p |\hat{K}_{ba}| \varphi_{q} \rangle.
\end{split}
\end{equation}
This procedure establishes a system of linear, coupled one-particle Schr\"{o}dinger-like equations in  Eq. (\ref{TDCIS}) for the orbitals $|\chi_a(t)\rangle$ with initial condition $|\chi_a(t_0)\rangle=0$.
To calculate photoelectron spectra, we need to evaluate with the transition amplitude between a modified Volkov state and the outgoing multi-channel wave packet, $\braket{\chi_{\mathbf{k},a}(t)}{\chi_a(t)}$, at a time, $t$, long after all interactions have ceased. 
The modified Volkov state in length gauge, $|\chi_{\mathbf{k},a}(t)\rangle$, with momentum $\mathbf{k}$ includes an additional phase-factor for a hole at the orbital $|\varphi_a\rangle$, and it satisfies the equation
\begin{align}
\label{modified_Volkov}
i\frac{\partial}{\partial t}|\chi_{\mathbf{k},a}(t)\rangle&=\left(-\frac{1}{2}\hat{\nabla}^2-\varepsilon_a-E(t)\hat{z}\right)|\chi_{\mathbf{k},a}(t)\rangle\nonumber \\ & \equiv\hat{H}_a(t)|\chi_{\mathbf{k},a}(t)\rangle,
\end{align}
where $\hat{H}_a(t)$ represents the modified Volkov Hamiltonian for the electron moving in the external laser field with a hole fixed at the orbital $|\varphi_a\rangle$.
To overcome the difficulty of the calculation with a large box, we adapt the time-dependent surface flux (t-SURFF) \cite{TaoNJP2012} method to the multi-channel TDCIS formalism. First, we define the overlap from a large distance $R_c$ to infinity between the modified Volkov state and the wave packet in a given channel, a, by a stepfunction:
\begin{align}
\label{transition_amplitude}
A_{\mathbf{k},a}(R_c,t)&\equiv \braOket{\chi_{\mathbf{k},a}(t)}{\theta(R_c)}{\chi_a(t)}
\nonumber \\ &=\int_{|\mathbf{r}|>R_c}d^{(3)}r\mbox{ }\chi_{\mathbf{k},a}^{*}(r,t)\chi_a(r,t),
\end{align}
that converges to the transition amplitude after some sufficiently large time $T_c$. 
Eq. (\ref{transition_amplitude}) can also be written as
\begin{align}
\label{transition_amplitude2}
 &A_{\mathbf{k},a}(R_c,T_c)=\int_{t_0}^{T_c} dt \frac{d}{dt} \langle\chi_{\mathbf{k},a}(t)|\theta(R_c)|\chi_{a}(t) \rangle
\nonumber  \\ &=\int_{t_0}^{T_c} dt  \Big[\Big(\frac{d}{dt} \langle\chi_{\mathbf{k},a}(t)|\Big) \theta(R_c)|\chi_{a}(t)\rangle 
\nonumber \\  &+ \langle\chi_{\mathbf{k},a}(t)| \theta(R_c)\frac{d}{dt}|\chi_{a}(t)\rangle\Big].
\end{align}
If we neglect correlation effects and ionic potential outside $R_c$, then the time-dependent wave packet $|\chi_a(t)\rangle$ follows the equation of motion
\begin{equation}
\label{TDCIS_asymp}
i\frac{\partial}{\partial t}|\chi_a(t)\rangle = \hat{H}_a(t)|\chi_a(t)\rangle+E(t)\sum_{b}z_{ba}|\chi_b(t)\rangle.
\end{equation}
In addition to the modified Volkov Hamiltonian, there is other term which makes different channels coupled to each other by laser field. With Eq. (\ref{modified_Volkov}), (\ref{transition_amplitude2}) and (\ref{TDCIS_asymp}), we can convert $A_{\mathbf{k},a}(R_c,T_c)$ from the spatial integration at $T_c$ to the temporal integration:
\begin{align}
\label{tsurff}
   &A_{\mathbf{k},a}(R_c,T_c) = i\int_{t_0}^{T_c} dt \Big[\langle\chi_{\mathbf{k},a}(t)| \hat{H}_a(t)\theta(R_c)|\chi_{a}(t)\rangle 
\nonumber \\ &
- \sum_{b}\langle\chi_{\mathbf{k},a}(t)| \theta(R_c) \Big(\hat{H}_a(t)\delta_{ab}+E(t) z_{ab}\Big)|\chi_{b}(t)\rangle\Big]
\nonumber \\ & 
                           = i\int_{t_0}^{T_c} dt \langle\chi_{\mathbf{k},a}(t)| [\hat{H}_a(t),\theta(R_c)] |\chi_{a}(t)\rangle 
\nonumber \\ &
- i \sum_{b}z_{ab}\int_{t_0}^{T_c} dt E(t)\langle\chi_{\mathbf{k},a}(t)| \theta(R_c) |\chi_{b}(t)\rangle 
\nonumber \\ &                           
= -\int_{t_0}^{T_c}dt J_{\mathbf{k},a}(R_c,t)-\int_{t_0}^{T_c}dt K_{\mathbf{k},a}(R_c,t).
\end{align}
We get two terms and the first term is the time integration of the flux 
\begin{align}
J_{\mathbf{k},a}(R_c,t)&=\frac{1}{2i}\Big[ -\chi_{\mathbf{k},a}^{*}(r,t)\partial_r \chi_a(r,t)
\nonumber \\ &
+\chi_a(r,t)\partial_r\chi_{\mathbf{k},a}^{*}(r,t) \Big]\Bigg|_{R_c}
\end{align}
through the boundary $R_c$ from $t_0$ to $T_c$ as indicated in \cite{TaoNJP2012,Karamatskou}. 
Compared with the previous work \cite{Karamatskou}, the second term is new and its integrand
\begin{equation}
	K_{\mathbf{k},a}(R_c,t)=i\sum_{b}z_{ab}E(t)\braOket{\chi_{\mathbf{k},b}(t)}{\theta(R_c)}{\chi_b(t)}e^{-i(\varepsilon_a-\varepsilon_b) t}
\end{equation}	 
  represents the channel-coupling of the TDCIS via laser field after the electronic wave packets pass though $R_c$. 
This contribution is missing in the integration of the flux at $R_c$, so this channel-coupling term can be viewed as an external source from other channels as the states of the ion makes transition.
In other words, if the field-driven transition between two different ionic states plays the role in the physical process, this channel-coupling term cannot be neglected.
This term is especially important if the photoelectron spectrum is measured in coincidence with parent ions.
The t-SURFF integral equation
\begin{equation}
 A_{\mathbf{k},a}(R_c,T_c)=-\int_{t_0}^{T_c}dt J_{\mathbf{k},a}(R_c,t)-\int_{t_0}^{T_c}dt K_{\mathbf{k},a}(R_c,t) 
\end{equation}
is numerically evaluated with $\chi_a(r,t)$ determined by TDCIS under the initial condition $A_{\mathbf{k},a}(R_c,t_0)=0$. 
Finally, the momentum spectrum $\sigma_{\mathbf{k},a}(\mathbf{k})$ and energy spectrum $\sigma_{E,a}(E)$ for the channel $a$ are given by
\begin{equation}
\sigma_{\mathbf{k},a}(\mathbf{k})\equiv|A_{\mathbf{k},a}(R_c,T_c)|^2
\end{equation}
\begin{equation}
\sigma_{E,a}(E)\equiv\sum_{|\mathbf{k}|=\sqrt{2E}}\frac{\sigma_{\mathbf{k}}(\mathbf{k})}{|\mathbf{k}|}.
\end{equation}
The numerical results presented in the main text are obtained using an 1D effective central potential and effective electron-electron repulsive potential
\begin{equation}
\begin{split}
	V_{\text{eff}}(z)     &=\frac{Z_\text{eff}}{\sqrt{z^2+z_c^2}} \\
	V_\text{ee}(z_1,z_2)&=\frac{Z_\text{ee}}{\sqrt{(z_1-z_2)^2+z_e^2}}
\end{split}
\end{equation}	
 with parameters $Z_\text{eff}$,$Z_\text{ee}$,$z_c$, and $z_e$ that reproduce the experimental ionization energies of the 2s and 2p orbitals in neon. We considered two models: First, the parameters of the effective potential were chosen such that the electron-electron interaction was zero, which corresponds to the independent particle approximation (IPA), by parametrization as $Z_\text{ee}=0$, $Z_\text{eff}=1.795$, and $z_c=0.7$. Second, we studied the correlated TDCIS model, parametrized by $Z_\text{ee}=1$, $Z_\text{eff}=9$, and $z_c=0.755$. The IPA and TDCIS agree remarkably well for our field parameters with only slightly different extracted phase parameters for the proposed method that depend on the detailed correlated interactions.


\begin{acknowledgments}
We thank Oliver M\"ucke, Thomas Pfeifer, Reinhard D\"orner and Eva Lindroth for stimulating discussions. 
J.M.D. acknowledges support from the Swedish Research Council, 
Grant No.  2013-344 and 2014-3724. 
\end{acknowledgments}



%
%

%
%
%


\end{document}